\documentclass[
reprint,
amsmath,
amssymb,
aps,
pra,
]{revtex4-1}

\usepackage{graphicx}
\usepackage{dcolumn}
\usepackage{bm}

\begin{document}

\title{ 
Stationary Entangled Radiation from Micromechanical Motion}
\author{S.~Barzanjeh}
\email{shabir.barzanjeh@ist.ac.at}
\author{E.~S.~Redchenko}
\author{M.~Peruzzo}
\author{M.~Wulf}
\author{D.~P.~Lewis}
\author{G.~Arnold}
\author{J.~M.~Fink}
\email{jfink@ist.ac.at}
\affiliation{Institute of Science and Technology Austria, 3400 Klosterneuburg, Austria}

\date{\today}

\begin{abstract}
Mechanical systems facilitate the development of a new generation of hybrid quantum technology comprising electrical, optical, atomic and acoustic degrees of freedom \cite{Aspelmeyer2014}. Entanglement is the essential resource that defines this new paradigm of quantum enabled devices. 
Continuous variable~(CV) entangled fields, known as Einstein-Podolsky-Rosen (EPR) states, are spatially separated two-mode squeezed states that can be used to implement quantum teleportation and quantum communication 
~\cite{
Braunstein2005}. In the optical domain, EPR states are typically generated using nondegenerate optical amplifiers \cite{
Andersen2016} and
at microwave frequencies Josephson circuits can serve as a nonlinear medium 
\cite{Eichler2011a, Flurin2012, Menzel2012}.
It is an outstanding goal to deterministically generate and distribute 
entangled states with a mechanical oscillator. 
Here we observe stationary emission of path-entangled microwave radiation from a parametrically driven 30 micrometer long silicon nanostring oscillator, squeezing the joint field operators of two thermal modes by 3.40(37)~dB below the vacuum level. 
This mechanical system correlates up to  50 photons/s/Hz 
giving rise to a quantum discord that is robust with respect to microwave noise \cite{Weedbrook2016}.
Such generalized quantum correlations of separable states are important for quantum enhanced detection \cite{Zhang2015} and provide direct evidence for the non-classical nature of the mechanical oscillator without directly measuring its state \cite{Krisnanda2017}. This noninvasive measurement scheme allows to infer information about otherwise inaccessible objects with potential implications in sensing, open system dynamics and fundamental tests of quantum gravity. In the near future, similar on-chip devices can be used to entangle subsystems on vastly different energy scales such as microwave and optical photons. 
\end{abstract}
\maketitle

Radiation pressure and back-action can give rise to entanglement and squeezing between electromagnetic radiation and a mechanical resonator \cite{Fabre1994,Mancini1994}. Recent experiments have demonstrated single mode squeezed states of mechanical motion \cite{Wollman2015,Lecocq2015a}
and radiation fields at both optical \cite{Safavi-Naeini2013b,Purdy2013b,Sudhir2017} and microwave~\cite{Korppi2017a} frequencies. Very recently, entanglement between photons and a mechanical oscillator~\cite{Palomaki2013b} and between two mechanical oscillators~\cite{
Riedinger2018,OckeloenKorppi2018} have been realized.  
Our results confirm the prediction that massive mechanical objects can deterministically produce path-entangled radiation~\cite{
Paternostro2007,
Genes2008b, 
Barzanjeh2011, Wang2013, Tian2013}. 
Using a highly versatile silicon-on-insulator electromechanical platform that is compatible with on-chip photonic \cite{Safavi-Naeini2013b, Pitanti2015} and phononic~\cite{Kalaee2018} crystal cavities, we demonstrate the generation of stationary entangled states between the propagating output fields of two microwave resonators separated by one meter in a millikelvin environment. Facilitated by the brightness of the mechanical squeezer, we prove the robustness of quantum discord in the presence of noise in the microwave domain. 

\begin{figure}[t!]
\centering
\includegraphics[width=1\columnwidth]{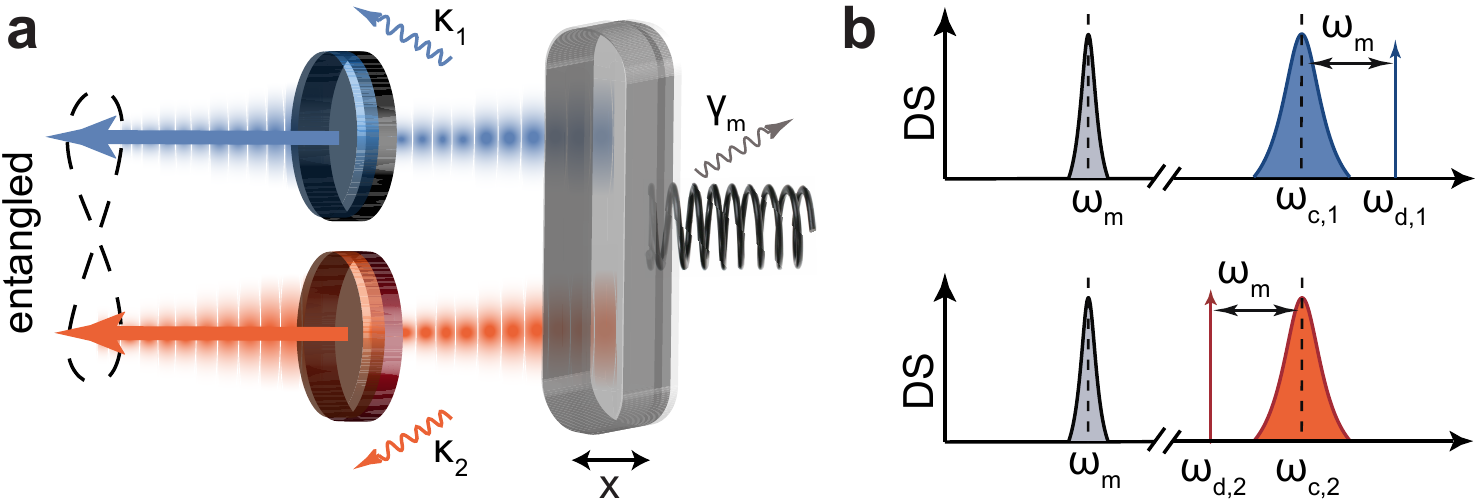}
\caption{\textbf{Schematic representation.} 
\textbf{a}, Two microwave cavities with a shared spring-loaded mirror generate two entangled output fields. $x$ indicates the mirror displacement, $\gamma_m$ denotes the mechanical loss rate and $\kappa_{1,2}$ the cavity loss rates. 
\textbf{b}, Density of states (DS) of the mechanical and optical oscillators with resonance frequencies $\omega_m$ and $\omega_{c,1,2}$ driven by two coherent tones (arrows) at $\omega_{d,1,2}$ on the blue and red detuned side of the two cavity resonances respectively.} \label{Fig1}
\end{figure}

We consider a three-mode electromechanical system in which two microwave cavities with resonance frequencies~$\omega_{c,i}$ and total damping rates~$\kappa_i$ with $i=1,2$ are 
coupled to a 
vibrational mirror mode 
with resonance frequency~$\omega_{m}$ and intrinsic damping rate~$\gamma_m$ as schematically shown in Fig.~\ref{Fig1}a. The electromagnetic field of the microwave resonators exerts radiation pressure on the mechanical resonator. In return, the vibration of the mechanical resonator mediates a retarded interaction between the microwave modes. In the presence of two strong microwave pumps with frequencies $\omega_{d,i}=\omega_{c,i}\pm\omega_m$ as indicated in Fig.~\ref{Fig1}b we can linearize the system and describe the physics in reference frames rotating at the frequencies~$\omega_{c,i}$ and~$\omega_{m}$ with the Hamiltonian
\begin{equation}\label{hamiltonian1}
H=G_1 (b^{\dagger} c_1^{\dagger}+c_1 b)+G_2 (b c_2^{\dagger}+c_2 b^{\dagger}),
\end{equation}
where $\hbar=1$, $c_i$ and $b$ are the annihilation operators for the cavity $i$
and the mechanical oscillator, and $G_i=g_{0,i}\sqrt{n_i}$ and $g_{0,i}$ are the effective and vacuum
electromechanical coupling rates between the mechanical mode and cavity $i$, respectively. $n_i$  is the number of photons in the resonator $i$ due to the drive with detuning~$\Delta_{1(2)}=\omega_{c,1(2)}-\omega_{d,1(2)}=\mp \omega_m$.
Here we assume the regime of fast mechanical oscillations, $\omega_m\gg \{\kappa_i,G_i\}$ 
which allows us to neglect the fast oscillating terms at~$\pm 2\omega_m$. 
The first term in Eq.~\ref{hamiltonian1} describes a parametric down-conversion interaction that is responsible for entangling the microwave resonator 1 with the mechanical oscillator. The second term describes a beam-splitter interaction between the mechanical resonator and the microwave resonator 2, exchanging the state of the electromagnetic and mechanical modes.
If the electromechanical coupling rate $4 G_i^2/\kappa_i$ exceeds the decoherence rate of the
mechanical resonator~$\gamma_m \bar{n}_m$, with the mechanical bath occupancy $\bar{n}_m=[e^{\hbar \omega_m/k_B T_b}-1]^{-1}$, $k_B$ is Boltzmann's constant and $T_b$ is the device temperature, the output of both microwave resonators is mapped into a two-mode squeezed thermal state \cite{Barzanjeh2011}, which can also be understood in the context of reservoir engineering \cite{Wang2013}.

\begin{figure*}[t!]
\centering
\includegraphics[width=2\columnwidth]{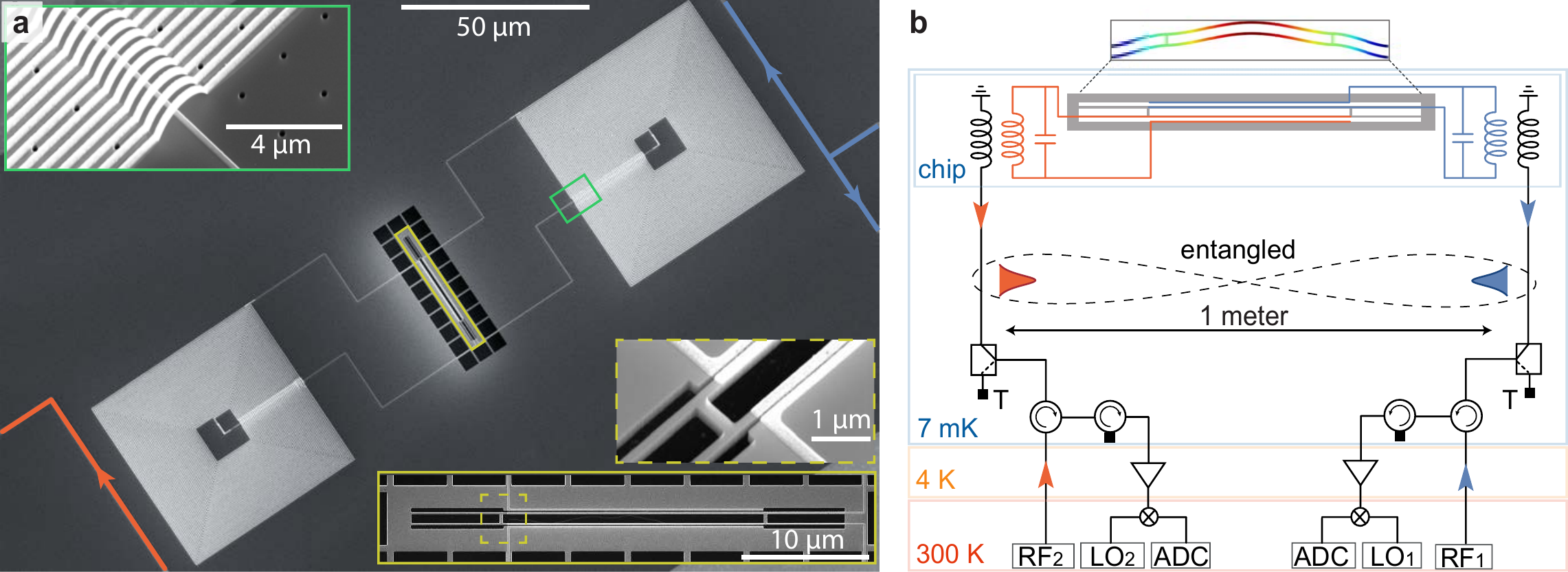}
\caption{\textbf{Experimental realization.} 
\textbf{a}, Scanning electron microscopy image of the microchip device composed of aluminum wires (bright gray) on silicon (dark gray). The insets depict details of the high impedance coil resonators inductively coupled to two coplanar waveguides via the feed-lines (blue and red), and the metalized and mechanically compliant double silicon nanobeam. The shown chip area forms a suspended 220 nm thick silicon membrane that is released in a hydrofloric acid vapor etch process. The etched squares surrounding the nanobeam reduce the effect of membrane buckling during thermal cycling. 
\textbf{b}, Circuit diagram of the experimental setup. Microwave drive tones RF$_{1,2}$ are filtered and attenuated before they enter the microchip, which is thermalized at 7~mK. The modulated and reflected tones are amplified, down-converted with a mixer and a local oscillator (LO$_{1,2}$) and digitized (ADC) simultaneously and independently for both channels. The sample is connected with a low loss printed circuit board and a pair of 50~cm long copper coaxial cables to two latching microwave switches which are used to select between the sample outputs and a temperature $T$ variable 50~$\Omega$ load (black squares) for the system calibration. The inset shows the simulated displacement of the in-plane flexural mode of the nanobeam used in the experiment, where color indicates relative displacement.} \label{Fig2}
\end{figure*}

We experimentally realize the described entanglement generation scheme in a hybrid dielectric-superconducting electromechanical system. The circuit, shown in~Fig.~\ref{Fig2}a consists of a metalized silicon nanobeam resonator who's vibrational in-plane mode at $\omega_m/2\pi=2.81$~MHz with an intrinsic damping rate of $\gamma_m/2\pi=6$~Hz and a bath occupation $\bar{n}_m=60$, is capacitively coupled to two high impedance superconducting coil resonators at $(\omega_{c,1},\omega_{c,2})/2\pi=(10.17, 12.13)$~GHz with energy decay rates~$(\kappa_{1},\kappa_{2})/2\pi=(0.52, 0.48)$~MHz and waveguide coupling ratios $(\eta_1,\eta_2)=(0.76,0.67)$. The strong vacuum electromechanical coupling strengths for this three-mode electromechanical system of $(g_{0,1},g_{0,2})/2\pi=(152,170)$~Hz are achieved by suspending and substantially miniaturizing the geometric inductors, see Appendix A for more details.
For the measurements shown below, the microwave resonator 1 (2) is driven from the blue (red) sideband with the coherent drive power $(P_b,P_r)=(-87.1,-84.4)$ dBm at the device input, corresponding to the single cavity cooperativites $C_i=4G_i^2/(\kappa_i \gamma_m)$ of~$(C_{1},C_{2})=(67.0,113.3)$.

The output of each resonator passes through two different measurement lines as shown in~Fig.~\ref{Fig2}b. After amplification the signals are filtered and down-converted to an intermediate frequency of 2~MHz and digitized with a sampling rate of 10~MHz using an 8 bit analog-to-digital converter. The reflected pumps are cancelled to avoid any amplifier compression. The FFT based digital down conversion process extracts the quadrature voltages $I_{i}$ and $Q_{i}$ for each channel. The measured quadrature voltages are converted to the unitless quadrature variables $X_i := (d_i+d_i^\dagger)/\sqrt{2}=I_i/\sqrt{\zeta_i}$ and $P_i := (d_i-d_i^\dagger)/\sqrt{2}\, \mathrm{i} =Q_i/\sqrt{\zeta_i}$ with the scaling factors $\zeta_i=  \mathcal{G}_i R B\hbar\omega_{c,i}$, where $d_i$ is the propagating resonator output mode and $\mathcal{G}_i$ is the total system gain of the output channel $i$, $B=100$~Hz is the digitally chosen measurement bandwidth, and $R=50\,\Omega$ the input impedance of the ADCs. We calibrate the system gain $(\mathcal{G}_1,\mathcal{G}_2)=(83.20(06), 79.99(08))$ dB and system noise $(n_{\mathrm{add,1}},n_{\mathrm{add,2}})=(8.3(1),11.5(2))$ of both measurement channels as described in Appendix B. We use these values for all following measurements, which locates our effective points of signal detection 0.5~m from the resonator outputs and approximately 1~m apart from each other as shown in Fig.~\ref{Fig2}b. Beyond this point, the generated states are exposed to amplifier noise, additional losses and a high temperature thermal bath.

\begin{figure*}[t!]
\centering
\includegraphics[width=1.8\columnwidth]{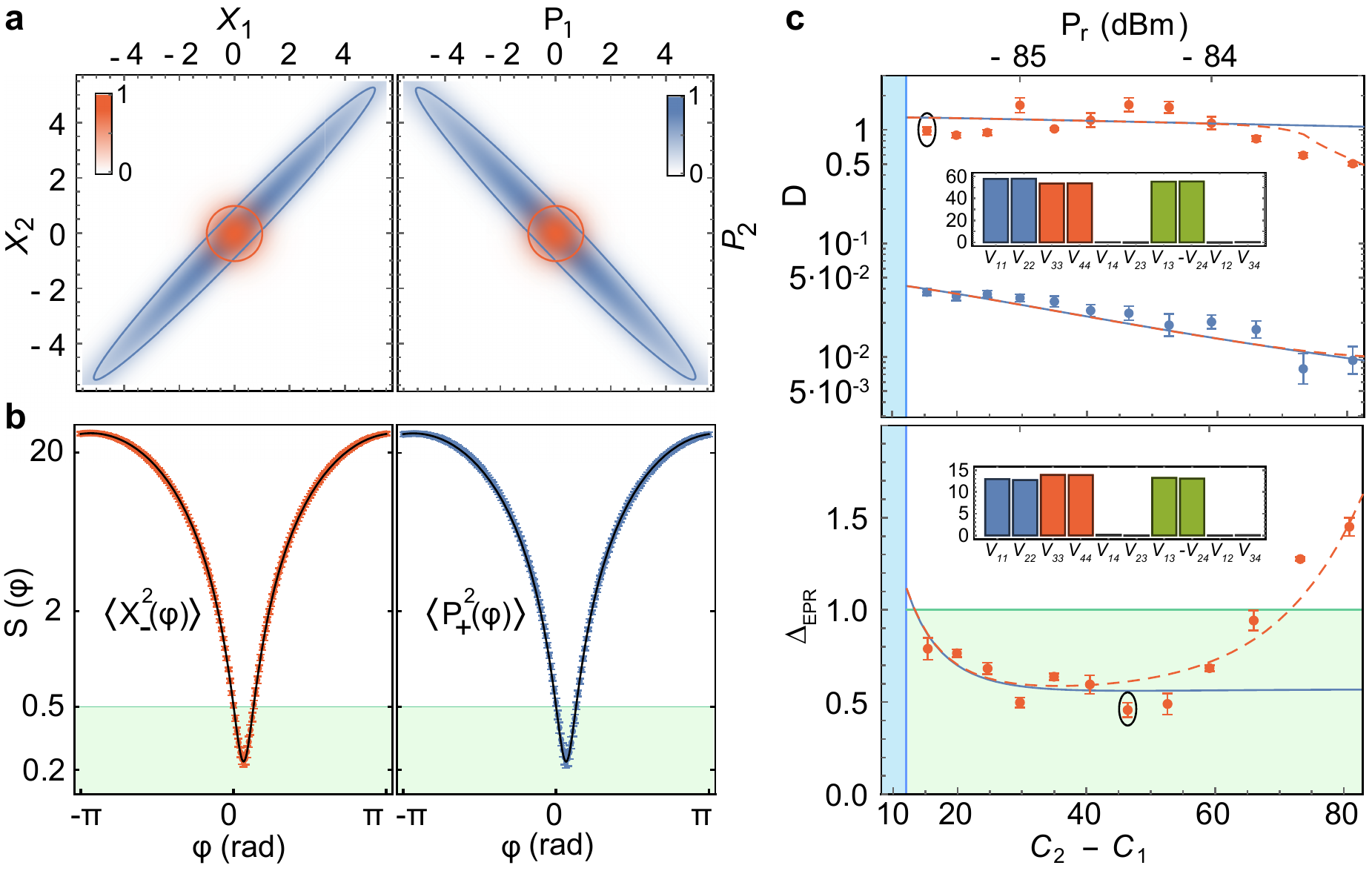}
\caption{
\textbf{Experimental results.} 
\textbf{a}, The reconstructed Wigner function Eq.~\ref{wigner} of the squeezed state (blue) in comparison to the ideal vacuum state (red) for two nonlocal quadrature pairs, where the other two variables are integrated out. Solid lines indicate a drop by $1/e$ of the maximum value. 
\textbf{b}, Measured variance of the EPR basis states $X_{-}(\varphi)$ and $P_{+}(\varphi)$ as a function of the detector angle $\varphi$ of channel 1 and fit to Eq.~\ref{squeezing} (solid lines). The green area shows the region of vacuum squeezing and the error bars indicate the standard deviation of the mean for 5 independent measurements of all 4 quadratures with 43200 values each. 
\textbf{c}, The measured Duan non-separability measure $\Delta_\mathrm{EPR}$ and quantum discord $D$ as a function of the pump power $P_r$ of the red detuned drive at the device input and the resulting cooperativity difference. Quantum discord is shown with (red) and without (blue) subtracting the calibrated system noise. The data and error bars are obtained identical to the ones in panel b. Theory is shown without (solid) and with pump induced noise (dashed). The green area indicates the region of entangled states and the blue area shows the unstable region where the theory breaks down and the measured squeezing values exceed the plot range. The insets show the relevant covariance matrix elements for the measurements with maximum squeezing and the maximum output photon number respectively (encircled).} 
\label{Fig4}
\end{figure*}

The non-classicality of such Gaussian states can be fully characterized by the $4\times 4$ covariance matrix $\mathbf{V}$, a symmetric matrix with 10 independent elements. The diagonal elements are calculated from the variances of the scaled quadratures when the pumps are on and off, i.e.~$V_{ii}=\langle u_i^2\rangle_{\mathrm{on}}-\langle u_i^2\rangle_{\mathrm{off}}+\frac{1}{2}\mathrm{coth} \frac{\hbar \omega_i}{2 k_B T_i}$ where $ u\in \{X_1,P_1,X_2,P_2\}$, $\frac{1}{2} \mathrm{coth}\,\hbar \omega_i/(2 k_B T_i)\approx 0.5 $ is the input quantum noise at temperature $T_i$ and the brackets~$\langle... \rangle$ show an average over all $216000$ ($604800$) measurements when the pumps are on (off). The off-diagonal elements of the covariance matrix are specified by the covariances of the two modes, $V_{ij}=\langle u_iu_j+u_ju_i\rangle_{\mathrm{on}}/2$, which are zero when the pumps are turned off. 
The amount of two-mode squeezing is best visualized using the quasiprobability Wigner function 
\begin{equation}\label{wigner}
W(\bm\psi)=\frac{\exp{\big(-\frac{1}{2} [\bm\psi \cdot \mathbf{V}^{-1} \cdot \bm\psi^\dagger]}\big)}{\pi^2 \sqrt{\det{[\mathbf{V}]}} }
\end{equation} 
with the state vector $\bm\psi=(X_1,P_1,X_2,P_2)$.
Figure~\ref{Fig4}a shows the two relevant Wigner function projections of the measured covariance matrix in blue and the ideal vacuum state $\mathbf{V_\mathrm{vac}}=\mathrm{\mathbf{I}}/2$ in red. The $\{X_1,X_2\}$ and $\{P_1,P_2\}$ projections clearly show cross-quadrature two mode squeezing below the quantum limit in the diagonal directions. The raw data of the measured Gaussian phase space representation, i.e. the two variable histograms representing the probability distribution of all possible combinations of the measured quadratures~$\{X_1,P_1,X_2,P_2\}$ are shown in Appendix B.


To quantitatively access the amount of squeezing we define the EPR operator pair $X_{-}=(X_1(\varphi)-X_2)/\sqrt{2}$ and  $P_{+}=(P_1(\varphi)+P_2)/\sqrt{2}$ where $\varphi$ represents a rotation of the detector phase in channel 1 implemented in post-processing. For each rotation angle we evaluate the squeezing parameters $\langle X_{-}^2(\varphi) \rangle = (V_{11} + V_{33} - 2 V_{13})/2$ and $\langle P_{+}^2(\varphi) \rangle = (V_{22} + V_{44} + 2 V_{24})/2$, as shown in Fig.~\ref{Fig4}b. For one common optimal rotation angle we find that both operators are squeezed by 3.43(38)~dB and 3.36(37)~dB below the vacuum level. The phase dependence shows squeezing and anti-squeezing as expected for a two-mode squeezer applied to a thermal Gaussian state (solid lines)
\begin{equation}\label{squeezing}
S(\varphi) = (1+n_1+n_2) \big(\cosh{(2 r)}-\sinh{(2 r)} \cos{(\varphi)}\big)/2
\end{equation}
with the effective thermal photon inputs $n_1=1.43$ and $n_2=2.49$ fully constrained by the measured output photon numbers $V_{11}=12.83$ and $V_{33}=13.89$ and the fitted squeezing parameter $r=1.19$ in agreement with the measured amount of correlations $V_{13}=13.13$, see Appendix C for more details.

To verify the existence of entanglement between the two output modes we use the Duan criterion \cite{Duan2000}. 
The two-mode Gaussian state is entangled if the parameter $\Delta_\mathrm{EPR}:=\langle X^2_{-}\rangle+\langle P^2_{+}\rangle<1$ and it is in a vacuum state for $\Delta_\mathrm{EPR}=1$. In Fig.~\ref{Fig4}c we show measurements of $\Delta_\mathrm{EPR}$ for the optimal angle $\varphi$ as a function of the red detuned drive power $P_r$ and the calculated difference between the red and blue cooperativities $C_2-C_1$. 
For small $P_r$ the system is predicted to be unstable 
and the measured values exceed the plot range. Far from this instability region (shaded in blue) at a cooperativity difference of 46, the inseparability condition is fulfilled and we find a minimum of $\Delta_\mathrm{EPR}=0.46(04)$, clearly proving that the radiation emitted from the electromechanical device is entangled before leaving the millikelvin environment. 
Increasing the red detuned pump $P_r$ the parameter $\Delta_\mathrm{EPR}$ eventually goes above the vacuum limit and the state becomes separable. We assign this effect to pump power dependent excitation of two-level systems which populate the microwave cavity with uncorrelated noise photons, that can result in a degradation of quantum correlations. We carefully ruled out amplifier nonlinearities with detuned pump-on measurements which would lead to the opposite effect (lower $\Delta_\mathrm{EPR}$ at higher pump power).
The reported errors are the statistical error of the measured means, which exceed the statistical errors and the measured long term variation of the subtracted noise measurement (pump turned off), as well as the error of the calibration measurements. 

To quantitatively understand the power dependence of the EPR parameter $\Delta_\mathrm{EPR}$ shown in Fig.~\ref{Fig4}c we fit the data with a full theoretical model outlined in Appendix C, which also takes into account the detection bandwidth and filter function.
At small pump powers above the instability we find excellent agreement with theory (blue solid line)
based only on the independently verified device parameters reported above. 
We quantify the degree of mechanically generated entanglement with the logarithmic negativity representing an upper bound of the distillable entanglement. 
The maximum measured entanglement in our system is $E_N=1.07(18)$ and the useable distribution rate of entangled EPR pairs of 127 ebits/s (entangled bits per second) can be calculated using the entropy of formation $E_F=0.45$ and the bandwidth of the emitted radiation $\gamma_\mathrm{eff}/2\pi
=282$~Hz. 

The quantum discord $D$ generalizes the concept of quantum correlations to  separable states. 
Figure~\ref{Fig4}c shows the extracted $D$ both with and without the calibrated system noise subtraction, see Appendix C for details. Without the presence of noise the discord is relatively stable and peaks at $D\sim1.5$ for the drive power where the entanglement is maximal. But even when no amplifier noise is subtracted we measure a positive discord over the full range, a hallmark for the nonclassical origin of the measured correlations. The maximum amount of $D=0.037(2)$ is obtained close to the instability region at $C_2-C_1=15$ where the correlated output photon numbers  $V_{13}= 55.25$ and $V_{24}= -55.41$
are maximal, as shown in the inset. The results are in excellent agreement with theory over the full power range and show that in the presence of noise higher output photon numbers result in larger discord, even if the initial amount of squeezing is lower. 

It has recently been shown that entanglement between the output modes proves that the mediating macroscopic mechanical oscillator is a non-classical object that must have shared quantum correlations - in the form of discord - with the two microwave modes for a finite time before reaching the steady state \cite{Krisnanda2017}. In the future, such a noninvasive technique could be used to determine the nature of other inaccessible or difficult to control mediators like the gravitational field or sensitive biological systems. 
In the near term, the presented mechanical entangler offers a step forward to 
harness the capabilities of superconducting circuits at elevated temperatures. On one hand, an analogous microwave - optical implementation with a photonic crystal cavity could be used to optically distribute entangled states between cold superconducting nodes \cite{Higginbotham2018}. On the other hand, we find that our cold electromechanical system produces noise-resilient quantum field correlations that could find use in quantum-enhanced microwave sensing applications \cite{Barzanjeh2015}, potentially even at room temperature. 

\noindent\textbf{Acknowledgements} 
This work was supported by IST Austria, the IST Nanofabrication Facility, the EU's Horizon 2020 research and innovation program under grant agreement No 732894 (FET Proactive HOT) and the European Research Council under the grant agreement No 758053 (ERC StG QUNNECT). S.B. acknowledges support from the Marie Sklodowska Curie fellowship No 707438 (MSC-IF SUPEREOM). G.A. is a recipient of a DOC fellowship of the Austrian Academy of Sciences at the Institute of Science and Technology. We thank N. Kuntner and J. Jung for contributions to the digitizer software, M. Hennessey-Wesen for developing the thermal calibration source, and K. Fedorov, M. Kalaee, O. Painter, D. Vitali and M. Paternostro for fruitful discussions. 


\onecolumngrid
\appendix

\section{Microchip design}
The sample used for this study has been fabricated in the IST Austria nanofabrication facility according to the recipe outlined in Ref.~\cite{Dieterle2016} using commercial high resistivity smart cut silicon on insulator substrates. For the anhydrous HF gas release, which is the last step of the process, we used the Orbis Alpha system from Memsstar. The sample parameters stated in the main text were extracted from thermometry and cooling measurements similar to Ref.~\cite{Fink2016}.  Compared to our earlier results based on silicon nanobeams~\cite{Fink2016,Barzanjeh2017}, a number of improvements helped to  increase the vacuum coupling strength by factor of $\sim$5, reaching $(g_{0,1},g_{0,2})/2\pi=(152,170)$~Hz. Our results confirm that narrow dielectric nanostrings with small motional capacitance can be very efficiently and reliably coupled to microwave circuits, identical to the best coupled purely metallic drumhead double devices \cite{Lecocq2015a}, and about 50 times stronger than the best previous single mode nanostring devices \cite{Teufel2009,Rocheleau2010}.

\subsection{Microwave design}
The two microwave resonators are realized as compact LC circuits suspended on a 220~nm thick silicon membrane with a 3 micrometer wide vacuum gap to the handle wafer. The planar coil inductors with 45 and 40 turns each and a wire to wire separation of 300 nm are made from a 200~nm wide and 100~nm thick aluminum wire resulting in the inductances $(L_{c,1},L_{c,2})/2\pi=(68, 51)$~nH as obtained from an analytic modified Wheeler method. These inductors are connected to the 2 vacuum gap nanobeam capacitors \cite{Pitanti2015} to form the microwave resonators. The calculated inductances and the measured resonance frequencies of $(\omega_{c,1},\omega_{c,2})/2\pi=(10.17, 12.13)$~GHz determine the total circuit capacitances of $(C_{\mathrm{tot},1},C_{\mathrm{tot},2})=(3.6,3.4)$~fF corresponding to a total characteristic impedance of $\sim 4$~k$\Omega$. With the mechanically modulated nanobeam capacitance of $\sim$1~fF on each side (see below), the total circuit stray capacitances are well below 3~fF. 

\subsection{Mechanical design}
The mechanical oscillator is formed by a micro-machined silicon double beam and subsequent metal deposition, as shown in Fig.~\ref{FigA1}a. The double beam is sectioned in 3 areas with two connecting tethers which serve the purpose to (i) couple the two beams, (ii) separate the modes in frequency from each other and from any out of plane modes, and (iii) minimize clamping losses of the in-plane differential mode. The large rectangular membrane cutouts surrounding the nanobeam help to avoid buckling which can cause hybridization with low frequency out-of-plane membrane modes, which would lead to mode splitting and reduced electromechanical coupling. 

During a cooldown from 300~K to 7~mK, the different thermal expansion coefficients of aluminum and silicon can lead to a differential stress in addition to any stress present at room temperature. Careful finite element method (FEM) simulations shown in Fig.~\ref{FigA1}b indicate that the tensile stress in the aluminum layer (silicon stress is taken to be zero) is $\sim600$~MPa at millikelvin temperatures for our samples. For this stress the simulated mechanical resonance frequencies $\omega_\textrm{m,sim}=(2.85,4.12)$~MHz agree very well with the experimentally measured frequencies of the two fundamental in-plane modes $\omega_\textrm{m,exp}=(2.81,4.10)$~MHz, shown in the insets.

Although the zero point fluctuations of the two nanobeam modes $x_\mathrm{zpf}= (32, 33)$~fm are very similar, the common in-plane mode (mode 1) exhibits a significantly higher coupling strength because a larger part of the beam is displaced during an oscillation. This is also reflected in a larger effective mass of $m_\mathrm{eff}=(2.9, 1.9)$~pg of mode 1. Due to this reason, and despite the requirements for lower phase noise sources, the experiment was performed with the stronger coupled common in-plane mode 1.

\subsection{Electromechanical coupling}

\begin{figure*}[t!]
\centering
\includegraphics[width=0.9\columnwidth]{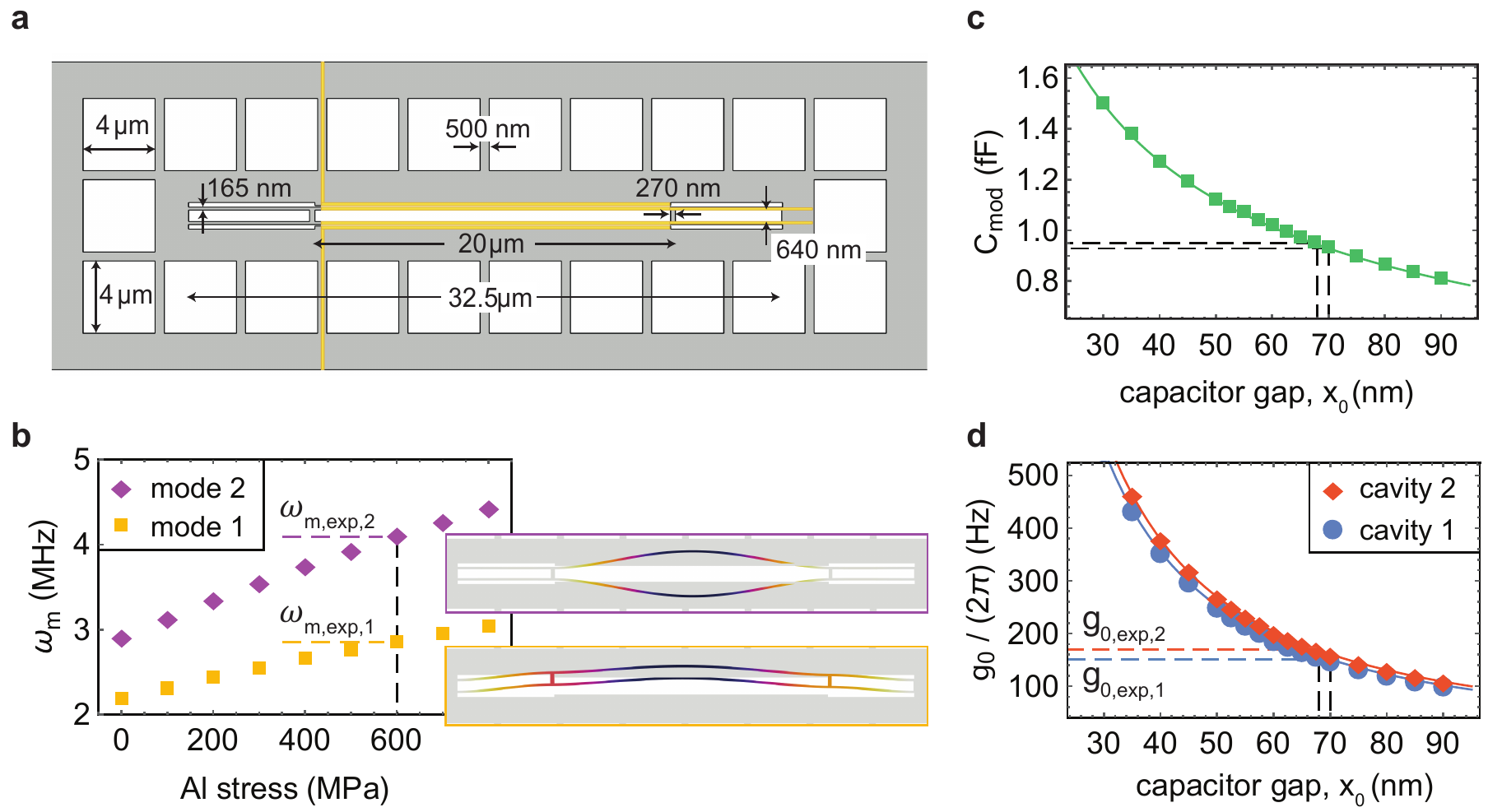}
\caption{
\textbf{FEM simulations.} 
\textbf{a}, Sample geometry. Important dimensions are indicated and the aluminum metallization is shown in yellow.
\textbf{b}, FEM simulated mechanical eigenfrequency of the two fundamental in-plane nanobeam modes versus tensile stress of the aluminum metallization of thickness $65$~nm. The insets show the displacement profile of the two modes.
\textbf{c}, Simulated modulated capacitance of a single beam as a function of the capacitor gap size. Solid line shows a fit with $C_\mathrm{mod}\propto x_0^{-0.6}$.
\textbf{d}, Simulated electromechanical coupling strength between the mechanical mode 1 and the two measured microwave resonators as a function of the capacitor gap size. Solid lines show fits with $g_0\propto x_0^{-1.5}$.} 
\label{FigA1}
\end{figure*}

The electromechanical vacuum coupling strength is given as 
\begin{equation}
g_{0}=-\beta \frac{\omega_{c}}{2 C_\textrm{mod}}\frac{dC_\mathrm{mod}}{dx} x_\mathrm{zpf},
\end{equation}
with the capacitive participation ratio $\beta=C_\textrm{mod}/C_\textrm{tot}$ and the amplitude coordinate $x$. The modulated capacitance $C_\mathrm{mod}$ strongly depends on the gap size of the capacitor. In Fig.~\ref{FigA1}c we show the result of an electrostatic simulation of the modulated capacitance of one side of the double beam vs. capacitor gap size $x_0$. As expected, the scaling $C_\mathrm{mod}\propto x_0^{-0.6}$ is somewhat weaker than that expected of a parallel plate capacitor. The derivative $dC_\mathrm{mod}/dx$ is calculated by a perturbation theory approach on the moving dielectric and metallic boundaries to the vacuum gap capacitor of the lower frequency common mode 1 \cite{Pitanti2015}. With the measured circuit resonance frequency and the calculated inductance, the effective vacuum coupling rate can then be calculated as a function of the capacitor gap size, which is shown in Fig.~\ref{FigA1}d. The experimentally determined values of $(g_{0,1},g_{0,2})/2\pi=(152,170)$~Hz correspond to effective gap sizes of $\sim$70 nm and $\sim$68 nm for the two electromechanically coupled resonators and for the same gap sizes we extract $C_\mathrm{mod}=(0.93, 0.95)$~fF. Due to the strong scaling $g_0\propto x_0^{-1.5}$ a reduction of the gap size as well as the circuit parasitic capacitance ($\beta\sim0.3$ for the current device) would lead to a significant further improvement in coupling strength. Capacitor gaps as small as 30~nm have been demonstrated in a similar fabrication process using shadow evaporated gold electrodes \cite{Pitanti2015}. 

\section{Quadrature measurements} 
\subsection{System noise calibration} 
We calibrate the system gain $\mathcal{G}_i$ and system noise $n_{\mathrm{add}, i}$ of both measurement channels 
by injecting a known amount of thermal noise using two temperature controlled $50\,\Omega$ loads \cite{Flurin2012,Ku2015}. The calibrators are attached to the measurement setup with two 5~cm long superconducting coaxial cables and a thin copper braid (for weak thermal anchoring to the mixing chamber plate) via two latching microwave switches (Radiall R573423600). By measuring the noise density in V$^2$/Hz at each temperature as shown in Fig.~\ref{Fig3}a, and fitting the obtained data with the expected scaling
\begin{equation}\label{noisedensity}
N_i=\zeta_i\big(1/2\, \mathrm{coth}[\hbar \omega_i/(2k_BT)]+n_{\mathrm{nadd,i}}\big)
\end{equation}
we accurately back out the gain $(\mathcal{G}_1,\mathcal{G}_2)=(83.20(06), 79.99(08))$ dB and the number of added noise photons $(n_{\mathrm{add,1}},n_{\mathrm{add,2}})=(8.3(1),11.5(2))$ for each output. The confidence values are taken from the standard error of the shown fit.

\subsection{Measured quadrature histograms} 
\begin{figure*}[t!]
\centering
\includegraphics[width=0.75\columnwidth]{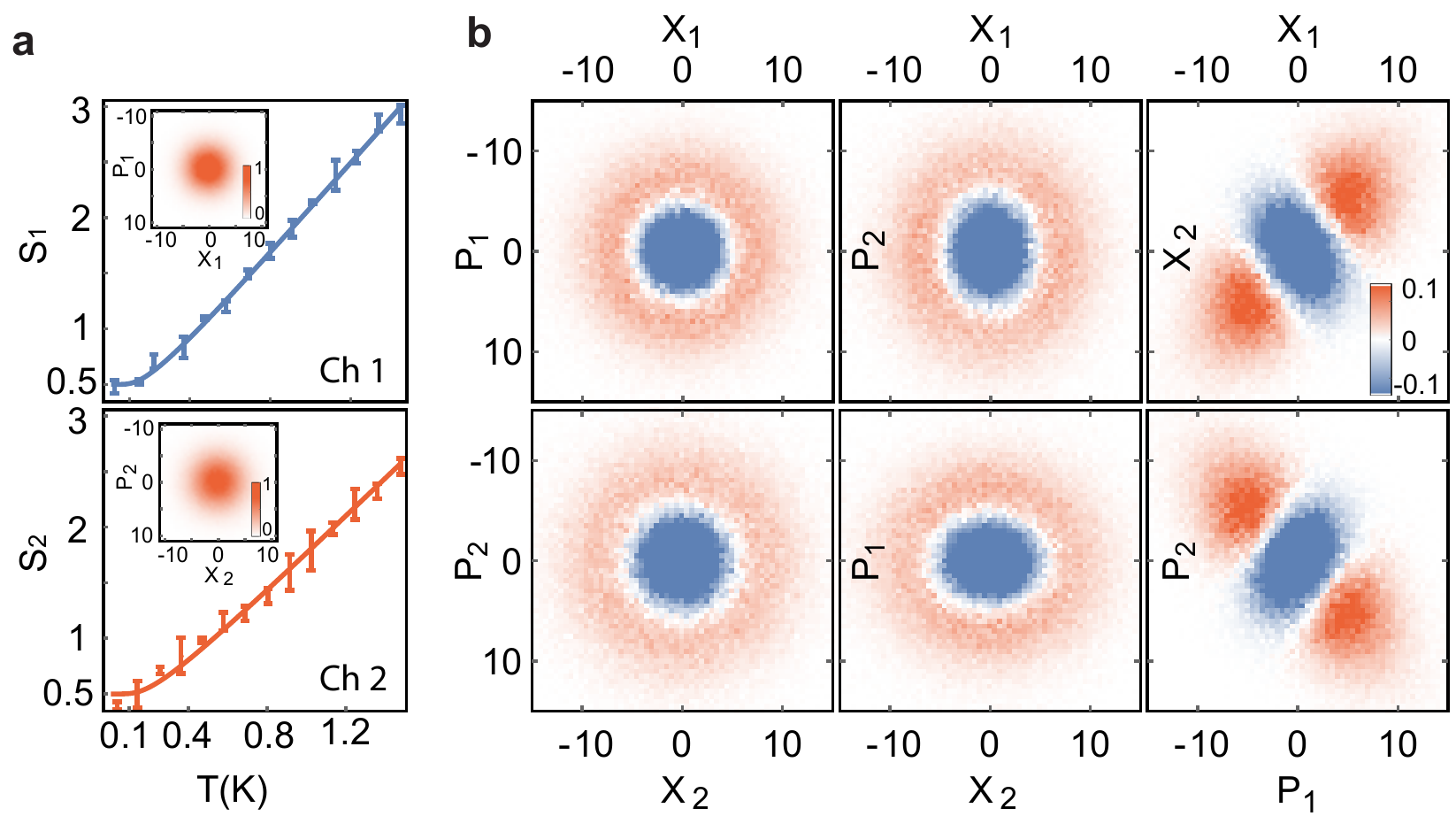}
\caption{
\textbf{Quadrature measurements.} 
\textbf{a}, System calibration of output channel 1 (2). The measured noise density in units of quanta S$_i=N_i/\zeta_i-n_{\mathrm{add},i}$ is shown as function of the temperature $T$ of the 50~$\Omega$ load. The error bars indicate the standard deviation of 3 measurements based on 28800 quadrature pairs each. The solid lines are fits to Eq.~\ref{noisedensity} in units of quanta, which yields the system gain and noise with the standard errors stated in the main text. 
The insets show the phase space distribution with the pump tones turned off. These two-variable quadrature histograms are based on 604800 measured quadrature pairs for each channel. 
\textbf{b}, The difference of the two-variable quadrature histograms with the pumps turned on and off for all quadrature pair combinations in units of quanta (as calibrated in panel A) based on 216000 value pairs from both channels.} 
\label{Fig3}
\end{figure*}

The generation of a two-mode squeezed thermal state can be verified intuitively in phase space by plotting the histograms representing the probability distribution of all possible combinations of the measured quadratures~$\{X_1,P_1,X_2,P_2\}$. 
We first measure the uncorrelated noise for each channel by performing a measurement with the microwave drives turned off. The result is shown in the insets of Fig.~\ref{Fig3}a indicating a thermal state with a variance corresponding to $n_{\mathrm{add,i}}+1/2$. 
In Fig.~\ref{Fig3}b we plot the 4 relevant quadrature histograms obtained when the drive tones are turned on (Rohde and Schwarz SMA100B-B711 and SMF100A) and after subtraction of the previously measured histogram with the drives turned off. The single-mode distributions $\{X_1,P_1\}$ and $\{X_2,P_2\}$ are both slightly broadened, indicating a phase-independent increase of the voltage fluctuations, which shows that the output of each resonator is amplified. For the chosen ideal rotation angle the same is true for the cross-mode distributions  $\{X_1,P_2\}$ and $\{X_2,P_1\}$. The slight stretching in the channel 2 direction is a result of the stronger red detuned pump power in channel 2, which results in a higher output photon number of 13.89 compared to the output photon number of 12.83 in channel 1. In stark contrast, in the histograms between the different outputs $\{X_1,X_2\}$ and $\{P_1,P_2\}$ the fluctuations increase along one diagonal axes and decrease in the other, indicating a strong correlation between the two spatially separated modes. 

\section{Theoretical model}
\subsection{Hamiltonian of double resonator electromechanics}

Our electromechanical system consists of a mechanical
resonator (MR) that is capacitively coupled to two
superconducting microwave resonators as depicted in Fig. 1 of the main text. These resonators' driving
fields are at radian frequencies $\omega _{\mathrm{d},j}=\omega _{c,j}-\Delta
_{0,j}$, where the $\Delta _{0,j}$ are the detunings from their resonant
frequencies $\omega _{c,j}$, with $j=1,2$. We include intrinsic losses for these
resonators with rates $\kappa _{j}^{\mathrm{in}}$, and use $\kappa _{j}^{%
\mathrm{ex}}$ to denote their input-port coupling rates. The Hamiltonian of
the coupled system in terms of annihilation and creation operators has been
studied in Ref. \cite{Barzanjeh2011}, and is given by
\begin{eqnarray}
\hat{H} =\hbar \omega _{m}\hat{b}^{\dagger }\hat{b}+\hbar \sum_{j=1,2}\Big[\omega _{c,j}\hat{a}_{j}^{\dagger }\hat{a}_{j}+ g_{0,j}(\hat{b}^{\dagger }+\hat{b})\hat{a}%
_{j}^{\dagger }\hat{a}_{j}+\mathrm{i} E_{j}(\hat{a}_{j}^{\dagger }e^{-\mathrm{i}%
\omega _{\mathrm{d},j}t}-\hat{a}_{j}e^{\mathrm{i}\omega _{\mathrm{d},j}t}) \Big].
\end{eqnarray}%
Here, $\hat{b}$ is the annihilation operator of the MR whose resonant
frequency is $\omega _{m}$, $\hat{a}_{j}$ is the annihilation operator for
resonator $j$ whose coupling rate to the MR is $g_{0,j}$. The microwave-driving
strength for resonator $j$ is $E_{j}=\sqrt{P_{j}\kappa _{j}^{\mathrm{ex}}/\hbar \omega _{%
\mathrm{d},j}}$, where $P_{j}$ is the amplitude of
the microwave driving field~\cite{Barzanjeh2011}. 

In the interaction picture with respect to $\hbar \omega _{\mathrm{d,1}}a_{%
\mathrm{1}}^{\dagger }a_{\mathrm{1}}+\hbar \omega _{\mathrm{d}%
,2}a_{2}^{\dagger }a_{2}$, and neglecting terms oscillating at $\pm 2\omega
_{\mathrm{d,j}}$, the system Hamiltonian reduces to
\begin{equation}
\hat{H}=\hbar \omega _{m}\hat{b}^{\dagger }\hat{b}+\hbar \sum_{j=1,2}\Big[\Delta _{0,j}+g_{0,j}(\hat{b}^{\dagger }+\hat{b})\Big]\hat{a}%
_{j}^{\dagger }\hat{a}_{j}+\hat{H}_{\mathrm{dri}},  \label{ham2}
\end{equation}%
where the Hamiltonian associated with the driving fields is $\hat{H}_{%
\mathrm{dri}}=\mathrm{i}\hbar \sum_{j=1,2}E_{j}(\hat{a}%
_{j}^{\dagger }-\hat{a}_{j})$.

We can linearize Hamiltonian~(\ref{ham2}) by expanding the resonator modes
around their steady-state field amplitudes, $\hat{c}_{j}=\hat{a}_{j}-\sqrt{%
n_{j}}$, where $n_{j}=|E_{j}|^{2}/(\kappa _{j}^{2}+\Delta _{j}^{2})\gg 1$ is
the mean number of intracavity photons induced by the microwave
pumps~\cite{Barzanjeh2011a}, the $\kappa _{j}=\kappa _{j}^{\mathrm{in}}+\kappa
_{j}^{\mathrm{ex}}$ are the total resonator decay rates, and the $\Delta _{j}$
are the effective resonator detunings. It is then convenient to move to the
interaction picture with respect to the free Hamiltonian, $\hbar \omega _{m}%
\hat{b}^{\dagger }\hat{b}+\hbar \sum_{j=1,2}\omega _{c,j}\hat{a}%
_{j}^{\dagger }\hat{a}_{j}$, where the linearized Hamiltonian becomes
\begin{equation}
\hat{H}=\hbar \sum_{j=1,2}G_{j}(\hat{b}\mathrm{e}^{-\mathrm{i}%
\omega _{m}t}+\hat{b}^{\dagger }\mathrm{e}^{\mathrm{i}\omega _{m}t})(\hat{c}%
_{j}^{\dagger }\mathrm{e}^{\mathrm{i}\Delta _{j}t}+\hat{c}_{j}\mathrm{e}^{-%
\mathrm{i}\Delta _{j}t}),  \label{ham3}
\end{equation}%
where $G_{j}=g_{0,j}\sqrt{n_{j}}$. By setting the effective resonator detunings
so that $\Delta _{1}=-\Delta _{2}=-\omega _{m}$ and neglecting the
terms rotating at $\pm 2\omega _{m}$, the above Hamiltonian reduces to
\begin{equation}
\hat{H}=\hbar G_{1}(\hat{c}_{1}\hat{b}+\hat{b}^{\dagger }\hat{c}%
_{1}^{\dagger })+\hbar G_{2}(\hat{c}_{2}\hat{b}^{\dagger }+%
\hat{b}\hat{c}_{2}^{\dagger }),  \label{hameff}
\end{equation}%
as specified in the main text.

The full quantum treatment of the system can be given in terms of the
quantum Langevin equations in which we add to the Heisenberg equations the
quantum noise acting on the mechanical resonator ($\hat b_{\text{in}} $
with damping rate $\gamma_m$), as well as the resonators' input fluctuations ($%
\hat c_{j,\mathrm{ex}}$, for $j=1,2$, with rates $\kappa_j^{%
\mathrm{ex}}$), plus the intrinsic losses of the resonator modes ($\hat c_{j,%
\mathrm{\ in}}$, for $j=1,2$, with loss rates $\kappa_j^{\mathrm{%
in}}$). These noises have the correlation functions
\begin{subequations}
\begin{align}
\langle \hat c_{j,\mathrm{ex}}(t) \hat c_{j,\mathrm{ex}}
^{\dagger}(t^{\prime})\rangle & = \langle \hat c_{j,\mathrm{ex}}
^{\dagger}(t) \hat c_{j,\mathrm{ex}}(t^{\prime})\rangle
+\delta(t-t^{\prime})=(\bar{n}_j^T+1)\delta(t-t^{\prime}), \\
\langle \hat c_{j,\mathrm{\ in}}(t) \hat c_{j,\mathrm{\ in}}
^{\dagger}(t^{\prime})\rangle & = \langle \hat c_{j,\mathrm{\ in}}
^{\dagger}(t) \hat c_{j,\mathrm{\ in}}(t^{\prime})\rangle
+\delta(t-t^{\prime})=(\bar{n}_j^{\mathrm{in}}+1)\delta(t-t^{\prime}), \\
\langle \hat b_{\mathrm{\ in}}(t) \hat b_{\mathrm{\ in}}
^{\dagger}(t^{\prime})\rangle & = \langle \hat b_{\mathrm{\ in}%
}^{\dagger}(t) \hat b_{\mathrm{\ in}}(t^{\prime})\rangle
+\delta(t-t^{\prime})=(\bar{n}_m+1)\delta(t-t^{\prime}),
\end{align}
where $\bar{n}_j^{\mathrm{\ in}}$, $\bar{n}_j$, and $\bar{n}_m$ are the
Planck-law thermal occupancies of each bath. The resulting Langevin
equations for the resonator modes and MR are
\end{subequations}
\begin{subequations}
\begin{align}  \label{qles2a}
\hat{\dot{c}}_1&=-\frac{\kappa_{1}}{2} \hat c_{1}-\mathrm{i}G_{1} \hat b+\sqrt{\kappa_{1}^{\mathrm{ex}}}\hat c_{
\mathrm{1,ex}}+\sqrt{\kappa_{1}^{\mathrm{in}}}\hat c_{\mathrm{
1,in}},\\
\hat{\dot{c}}_{2}&=-\frac{\kappa_{2}}{2} \hat c_{2}-\mathrm{i}G_{2} \hat
b^{\dagger}+\sqrt{\kappa_{2}^{\mathrm{ex}}}\hat c_{2,\mathrm{ex}}+\sqrt{%
\kappa_{2}^{\mathrm{in}}}\hat c_{2,\mathrm{in}}, \\
\hat{\dot{b}}&=-\frac{\gamma_m}{2} \hat b-\mathrm{i}G_{{1}} \hat c_{{1}}^{\dagger}-%
\mathrm{i}G_{\mathrm{2}} \hat c_{\mathrm{2}}+\sqrt{\gamma_m}\hat b_{\mathrm{
\ in}}.  \label{qles2c}
\end{align}
We can solve the above equations in the Fourier domain to obtain the
microwave resonators variables. By substituting the solutions of
Eqs.~(\ref{qles2a})--(\ref{qles2c}) into the corresponding input-output
formula for the resonators' variables, i.e., $\hat d_j\equiv \hat c_{j,\text{%
out}}= \sqrt{\kappa_j^{\mathrm{ex}}}\hat c_j-\hat c_{j,\mathrm{ex}}$, we
obtain
\end{subequations}
\begin{subequations}
\begin{align}  \label{qlessimplifya}
\hat d_{\mathrm{1}}(\omega)&=\alpha_1(\omega)\hat c_{\mathrm{1,ex}%
}+\alpha_{12}(\omega)\hat c_{2,\mathrm{ex}}^{\dagger}+\alpha_{1m}(\omega) \hat b^{\dagger}_{%
\mathrm{in}}+\alpha_{1in}(\omega)\hat c_{1,\mathrm{in}}+\alpha_{12in}(\omega) \hat c_{\mathrm{2,in}}^{\dagger}, \\
\hat d_{\mathrm{2}}(\omega)&=\alpha_2(\omega)\hat c_{\mathrm{2,ex}%
}+\alpha_{21}(\omega)\hat c_{1,\mathrm{ex}}^{\dagger}+\alpha_{2m}(\omega) \hat b_{%
\mathrm{in}}+\alpha_{2in}(\omega)\hat c_{2,\mathrm{in}}+\alpha_{21in}(\omega) \hat c_{\mathrm{1,in}}^{\dagger}, 
\label{qlessimplifyc}
\end{align}
\end{subequations}
where
\begin{subequations}  \label{coeffa}
\begin{align}
\alpha_1(\omega)&=-1+\frac{2\eta_1\big[\tilde{\omega}_2\tilde{\omega}_b+\mathcal{C}_2\big]}{\tilde{\omega}_1\mathcal{C}_2+\tilde{\omega}_2(\tilde{\omega}_1\tilde{\omega}_b-\mathcal{C}_1)}\\
\alpha_{12}(\omega)&=\frac{2\sqrt{\eta_1\eta_2\,\mathcal{C}_1\mathcal{C}_2}}{\tilde{\omega}_1\mathcal{C}_2+\tilde{\omega}_2(\tilde{\omega}_1\tilde{\omega}_b-\mathcal{C}_1)}\\
\alpha_{1m}(\omega)&=\frac{2\mathrm{i}\sqrt{\eta_1\,\mathcal{C}_1}\tilde{\omega}_2}{\tilde{\omega}_1\mathcal{C}_2+\tilde{\omega}_2(\tilde{\omega}_1\tilde{\omega}_b-\mathcal{C}_1)}\\
\alpha_{1in}(\omega)&=\frac{2\sqrt{\eta_1(1-\eta_1)}\,(\tilde{\omega}_2\tilde{\omega}_b+\mathcal{C}_2)}{\tilde{\omega}_1\mathcal{C}_2+\tilde{\omega}_2(\tilde{\omega}_1\tilde{\omega}_b-\mathcal{C}_1)}\\
\alpha_{12in}(\omega)&=\frac{2\sqrt{\eta_1(1-\eta_2)\,\mathcal{C}_1\mathcal{C}_2}}{\tilde{\omega}_1\mathcal{C}_2+\tilde{\omega}_2(\tilde{\omega}_1\tilde{\omega}_b-\mathcal{C}_1)}\\
\end{align}
\end{subequations}
and
\begin{subequations}  \label{coeffa1}
\begin{align}
\alpha_2(\omega)&=-1+\frac{2\eta_2\big[\tilde{\omega}_1\tilde{\omega}_b-\mathcal{C}_1\big]}{\tilde{\omega}_1\mathcal{C}_2+\tilde{\omega}_2(\tilde{\omega}_1\tilde{\omega}_b-\mathcal{C}_1)}\\
\alpha_{21}(\omega)&=-\frac{2\sqrt{\eta_1\eta_2\,\mathcal{C}_1\mathcal{C}_2}}{\tilde{\omega}_1\mathcal{C}_2+\tilde{\omega}_2(\tilde{\omega}_1\tilde{\omega}_b-\mathcal{C}_1)}\\
\alpha_{2m}(\omega)&=-\frac{2\mathrm{i}\sqrt{\eta_2\,\mathcal{C}_2}\tilde{\omega}_1}{\tilde{\omega}_1\mathcal{C}_2+\tilde{\omega}_2(\tilde{\omega}_1\tilde{\omega}_b-\mathcal{C}_1)}\\
\alpha_{2in}(\omega)&=\frac{2\sqrt{\eta_2(1-\eta_2)}\,(\tilde{\omega}_1\tilde{\omega}_b-\mathcal{C}_1)}{\tilde{\omega}_1\mathcal{C}_2+\tilde{\omega}_2(\tilde{\omega}_1\tilde{\omega}_b-\mathcal{C}_1)}\\
\alpha_{21in}(\omega)&=-\frac{2\sqrt{\eta_2(1-\eta_1)\,\mathcal{C}_1\mathcal{C}_2}}{\tilde{\omega}_1\mathcal{C}_2+\tilde{\omega}_2(\tilde{\omega}_1\tilde{\omega}_b-\mathcal{C}_1)}\\
\end{align}
\end{subequations}
with $\tilde{\omega}_j=1-\mathrm{i}\omega/\kappa_j$, $\tilde{\omega}_b=1-%
\mathrm{i}\omega/\gamma_m$, $\eta_i=\kappa_i^{\mathrm{ex}}/\kappa_i$,  and $\mathcal{C}_j =4 G^2_j/\kappa_j\gamma_m$. The
coefficients (\ref{coeffa})--(\ref{coeffa1}) become much simpler at $%
\omega\simeq 0 $, which corresponds to take a narrow frequency band around
each resonator resonance, viz.,

\begin{subequations}  \label{coeffa2}
\begin{align}
\alpha_1(\omega)&=-1+\frac{2\gamma_m\eta_1\big[1+\mathcal{C}_2\big]}{\gamma_{\mathrm{eff}}}\\
\alpha_{12}(\omega)&=\frac{2\gamma_m\sqrt{\eta_1\eta_2\,\mathcal{C}_1\mathcal{C}_2}}{\gamma_{\mathrm{eff}}}\\
\alpha_{1m}(\omega)&=\frac{2\mathrm{i}\gamma_m\sqrt{\eta_1\,\mathcal{C}_1}}{\gamma_{\mathrm{eff}}}\\
\alpha_{1in}(\omega)&=\frac{2\gamma_m\sqrt{\eta_1(1-\eta_1)}\,(1+\mathcal{C}_2)}{\gamma_{\mathrm{eff}}}\\
\alpha_{12in}(\omega)&=\frac{2\gamma_m\sqrt{\eta_1(1-\eta_2)\,\mathcal{C}_1\mathcal{C}_2}}{\gamma_{\mathrm{eff}}}\\
\end{align}
\end{subequations}
and
\begin{subequations}  \label{coeffa3}
\begin{align}
\alpha_2(\omega)&=-1+\frac{2\gamma_m\eta_2\big[1-\mathcal{C}_1\big]}{\gamma_{\mathrm{eff}}}\\
\alpha_{21}(\omega)&=-\frac{2\gamma_m\sqrt{\eta_1\eta_2\,\mathcal{C}_1\mathcal{C}_2}}{\gamma_{\mathrm{eff}}}\\
\alpha_{2m}(\omega)&=-\frac{2\mathrm{i}\gamma_m\sqrt{\eta_2\,\mathcal{C}_2}}{\gamma_{\mathrm{eff}}}\\
\alpha_{2in}(\omega)&=\frac{2\gamma_m\sqrt{\eta_2(1-\eta_2)}\,(1-\mathcal{C}_1)}{\gamma_{\mathrm{eff}}}\\
\alpha_{21in}(\omega)&=-\frac{2\gamma_m\sqrt{\eta_2(1-\eta_1)\,\mathcal{C}_1\mathcal{C}_2}}{\gamma_{\mathrm{eff}}}\\
\end{align}
\end{subequations}
with  $\gamma_{\mathrm{eff}}=\gamma_m(1+\mathcal{C}_2-\mathcal{C}_1)$ is the effective damping rate of the MR. Furthermore, when the internal losses are negligible, i.e., $\eta_j=1$, then we get $\alpha_{1in}=\alpha_{2in}=\alpha_{12in}=\alpha_{21in}=0 $, and
Eqs.~(\ref{qlessimplifya})--(\ref{qlessimplifyc}) reduce to the simple forms
\begin{subequations}
\begin{align}\label{simplea}
\hat d_{\mathrm{1}}&=\alpha_1\hat c_{\mathrm{1,ex}%
}+\alpha_{12}\hat c_{2,\mathrm{ex}}^{\dagger}+\alpha_{1m} \hat b^{\dagger}_{%
\mathrm{in}} \\
\hat d_{\mathrm{2}}&=\alpha_2\hat c_{\mathrm{2,ex}%
}+\alpha_{21}\hat c_{1,\mathrm{ex}}^{\dagger}+\alpha_{2m} \hat b_{%
\mathrm{in}},
\label{simpleb}
\end{align}
with coefficients given by
\end{subequations}
\begin{subequations}
\begin{align}  \label{papercoeffa}
\alpha_1&=-1+\frac{2\gamma_m\big[1+\mathcal{C}_2\big]}{\gamma_{\mathrm{eff}}}\\
\alpha_2&=-1+\frac{2\gamma_m\big[1-\mathcal{C}_1\big]}{\gamma_{\mathrm{eff}}}\\
\alpha_{12}&=-\alpha_{21}=\frac{2\gamma_m\sqrt{\mathcal{C}_1\mathcal{C}_2}}{\gamma_{\mathrm{eff}}}\\
\alpha_{1m}&=\frac{2\mathrm{i}\gamma_m\sqrt{\mathcal{C}_1}}{\gamma_{\mathrm{eff}}}\\
\alpha_{2m}&=-\frac{2\mathrm{i}\gamma_m\sqrt{\mathcal{C}_2}}{\gamma_{\mathrm{eff}}},
\label{papercoeffc}
\end{align}
\end{subequations}

These input-output relations preserve the bosonic commutation relations,
i.e., when the operators on the right in Eqs.~(\ref{simplea}) and (\ref{simpleb}%
) satisfy those commutation relations, we get $[\hat d_i,\hat
d_j^\dagger]=\delta_{i,j}$ and $[\hat d_i,\hat d_j]=[\hat d_i^\dagger,\hat
d_j^\dagger]=0$, for $i,j\in 1,2$.

The system is stable if the Routh-Hurwitz criterion is
satisfied. For $\mathcal{C} _{i}\gg 0$, this criterion reduces to the following
necessary and sufficient condition~\cite{Wang2013}:
\begin{equation*}
\kappa _{\text{2}}\,\mathcal{C}_{\text{2}}-\kappa _{\text{1}}\mathcal{C}_{\text{1}}>%
\tilde{\mathcal{C}}\,\text{max}\left\{ \kappa _{\text{2}}-\kappa _{\text{1}},\frac{%
\kappa _{\text{1}}^{2}-\kappa _{\text{2}}^{2}}{2\gamma _{m}+\kappa _{\text{1}%
}+\kappa _{\text{2}}}\right\} ,
\end{equation*}%
where $\tilde{\mathcal{C}}=\frac{\mathcal{C} _{\text{2}}}{1+\kappa _{\text{1}}/\kappa _{%
\text{2}}}+\frac{\mathcal{C} _{\text{1}}}{1+\kappa _{\text{2}}/\kappa _{\text{1}}}
$.

\subsection{Covariance matrix of a two-mode Gaussian state}

In order to quantify entanglement, we first determine the
covariance matrix~(CM) of our system in the frequency domain,
which can be expressed as
\begin{equation}
V_{ij}=\frac{1}{2}\langle
u_{i}u_{j}+u_{j}u_{i}\rangle ,  \label{cor1}
\end{equation}%
where
\begin{equation}
\mathbf{u}=[X_1,Y_1,X_2,Y_2]^{T},
\end{equation}%
and $X_{j}=(D_{j}+D_{j}^{\dagger })/\sqrt{2}$,
$Y_{j}=(D_{j}-D_{j}^{\dagger })/\mathrm{i}\sqrt{2}$ with
$j=1,2$. Note that the vacuum noise has variance $1/2$ in
these quadratures. Here we have defined the filtered output operators 
\begin{equation}
D_j(B)=\int_{-\infty}^{\infty}d\omega' f_j(\omega',B)d_j(\omega')
\end{equation}
where a filter function $f_j(\omega,B)$ with bandwidth $ B $ is applied on the output of the each resonator. 
Now, by using Eqs.~(\ref{qlessimplifya}), (\ref{qlessimplifyc}) and~(\ref{cor1}), we obtain
the CM for the quadratures of the resonators outputs,
which is given by the normal form
\begin{equation}
\mathbf{V}(\omega )=\left(
\begin{array}{cccc}
V_{11} & 0 & V_{13} & 0 \\
0 & V_{11} & 0 & -V_{13} \\
V_{13} & 0 & V_{33} & 0 \\
0 & -V_{13} & 0 & V_{33}%
\end{array}
\right) ,  \label{driftA}
\end{equation}
Note that Eq.~(\ref{driftA}) is the typical CM of a two-mode squeezed thermal state~\cite{Olivares2012,Pirandola2014} where the elements of the CM can be written in terms of photon numbers $n_i$, squeezing angle $\phi$ and squeezing parameter $r$, reads
\begin{subequations}
\begin{align}  \label{SQa}
V_{11}&=V_{22}=\frac{(1+n_1+n_2)\mathrm{cosh}(2r)+(n_1-n_2)}{2},\\
V_{33}&=V_{44}=\frac{(1+n_1+n_2)\mathrm{cosh}(2r)-(n_1-n_2)}{2},\\
V_{13}&=-V_{24}=\frac{(1+n_1+n_2)\mathrm{sinh}(2r)\,\mathrm{cos}\phi}{2},
\label{SQb}
\end{align}
\end{subequations}

when $n_i=0$ the Gaussian state is called two-mode squeezed vacuum. Squeezing in the two-mode squeezed thermal state can be determined by following expression
\begin{equation}
S(\phi)=V_{11}+V_{33}-2V_{13}=\frac{1}{2}(1+n_1+n_2)\Big(\mathrm{cosh}(2r)-\mathrm{sinh}(2r)\mathrm{cos}\phi\Big),
\end{equation}
For $\phi=0$ and $n_i=0$ we get $S(0)=e^{-2r}/2$.
 
\subsection{Logarithmic Negativity}
Here we quantify the amount of entanglement generated by our
microwave entanglement source using standard measures in quantum information
theory. In particular, we consider the log-negativity~\cite{Vidal2002, Plenio2005}, which is an upper bound to the number of distillable entanglement bits (ebits) generated
by the source. 

The log-negativity $E_{N}$ is given by~\cite{Vidal2002, Plenio2005}

\begin{equation}
E_{N}=\mathrm{max}[0,-\mathrm{\log }(2\zeta ^{-})],  \label{loga}
\end{equation}%
where $\zeta ^{-}$ is the smallest partially-transposed symplectic
eigenvalue of $\mathbf{V}(\omega )$, given by~\cite{Weedbrook2012}%
\begin{equation}
\zeta ^{-}=2^{-1/2}\left( V_{11}^{2}+V_{33}^{2}+2V_{13}^{2}-\sqrt{%
(V_{11}^{2}-V_{33}^{2})^{2}+4V_{13}^{2}(V_{11}+V_{33})^{2}}\right) ^{1/2}.
\end{equation}%

\subsection{Quantum correlations beyond entanglement: Quantum discord}

Our microwave source generates a Gaussian state which is mixed, as
one can easily check from the numerical values of its von Neumann entropy. It is therefore important to describe its quality in terms
of general quantum correlations beyond quantum entanglement. Thus we compute
here the quantum discord~\cite{Ollivier2001,Henderson2001} of the source $D($2$|1)$,
capturing the basic quantum correlations which are carried by the microwave
modes. 

Since our source emits a mixed Gaussian state which is a two-mode squeezed
thermal state, we can compute its (unrestricted) quantum discord using the
formulas of Ref.~\cite{Pirandola2014}. In particular, the CM in Eq.~(\ref{driftA})
can be expressed as
\begin{equation}
\mathbf{V}(\omega )=\left(
\begin{array}{cc}
(\tau b+\eta )\mathbf{I} & \sqrt{\tau (b^{2}-1)}\mathbf{Z} \\
\sqrt{\tau (b^{2}-1)}\mathbf{Z} & b\mathbf{I}%
\end{array}%
\right) ,~~%
\begin{array}{c}
\mathbf{I}\equiv \mathrm{diag}(1,1),~~ \\
\mathbf{Z}\equiv \mathrm{diag}(1,-1),%
\end{array}%
\end{equation}%
where%
\begin{equation}
b=V_{33},~~\tau =\frac{V_{13}^{2}}{V_{33}^{2}-1},~~\eta =V_{11}-\frac{%
V_{33}V_{13}^{2}}{V_{33}^{2}-1}.
\end{equation}%
Thus, we may write~\cite{Pirandola2014}%
\begin{eqnarray}
D(2|1) &=&h(b)-h(\nu _{-})-h(\nu _{+})+h(\tau +\eta ) \\
&=&h(V_{33})-h(\nu _{-})-h(\nu _{+})+h\left[ V_{11}+\frac{%
V_{13}^{2}(1-V_{33})}{V_{33}^{2}-1}\right] ,
\end{eqnarray}%
where $\nu _{-}$ and $\nu _{+}$ are the symplectic eigenvalues of $\mathbf{V}%
(\omega )$ and they are given by~\cite{Weedbrook2012}
\begin{equation}
\nu _{\pm }=2^{-1/2}\left( V_{11}^{2}+V_{33}^{2}-2V_{13}^{2}\pm \sqrt{%
(V_{11}^{2}-V_{33}^{2})^{2}-4V_{13}^{2}(V_{11}-V_{33})^{2}}\right) ^{1/2}.
\end{equation}%
where
\begin{equation}
h(x)\equiv \left( x+\frac{1}{2}\right) \mathrm{log}\left( x+\frac{1}{2}%
\right) -\left( x-\frac{1}{2}\right) \mathrm{log}\left( x-\frac{1}{2}%
\right).
\end{equation}
Note that the expression of the entropic function $h(x)$ is
that for vacuum noise equal to $1/2$. Our notation is different from that of
Ref.~\cite{Weedbrook2012}, where the vacuum noise is equal to $1$.
\subsection{Entropy of formation}
The effective number of ebits at the detectors input known is entropy of formation can be expressed in terms of the log-negativity defined in Eq. \ref{loga} \cite{Giedke2003, Flurin2012, Schneider2018}
\begin{equation}
E_f=\sigma_+\mathrm{log}_2\sigma_+-\sigma_-\mathrm{log}_2\sigma_-,
\end{equation}
where $\sigma_{\pm}=(\frac{1}{\sqrt{\theta}}\pm\sqrt{\theta})^2/4$ with $\theta=2^{-E_N}$. 

\end{document}